\documentclass[conference, a4paper]{IEEEtran}

\IEEEoverridecommandlockouts                              
\usepackage{units}
\usepackage{graphicx}
\usepackage{multirow}
\usepackage{amsmath,amssymb,latexsym}
\usepackage{rotating}
\usepackage{textcomp}
\usepackage{url}
\usepackage{hyperref}

\IEEEoverridecommandlockouts

\newcommand{\Ctrq}{C_\mathrm{fr}}

\newcommand{\Ctls}{C_\mathrm{tx}}
\newcommand{\Crls}{C_\mathrm{rx}}
\newcommand{\Nsleep}{N_\mathrm{slp}}
\newcommand{\Nsnooze}{N_\mathrm{snz}}
\newcommand{\Nwup}{n_\mathrm{wup}}
\newcommand{\Tslot}{T_\mathrm{slot}}
\newcommand{\Tsf}{T_\mathrm{sf}}
\newcommand{\Twc}{T_\mathrm{wc}}
\newcommand{\Tc}{T_\mathrm{c}}
\newcommand{\Tmin}{T_\mathrm{d}}

\newcommand{\tauc}{\tau_\mathrm{c}}
\newcommand{\taud}{\tau_\mathrm{d}}

\newcommand{\Lambdasf}{\Lambda_\mathrm{sf}}
\newcommand{\Lambdac} {\Lambda_\mathrm{c}}
\newcommand{\Ptx} {P_\mathrm{t}}
\newcommand{\Prx} {P_\mathrm{r}}
\newcommand{\Prz} {P_\mathrm{r0}}

\newcommand{\Etxd} {E_\mathrm{txd}}
\newcommand{\Erxd} {E_\mathrm{rxd}}
\newcommand{\Etxa} {E_\mathrm{txa}}
\newcommand{\Erxa} {E_\mathrm{rxa}}
\newcommand{\Etxe} {E_\mathrm{txe}}
\newcommand{\Erxe} {E_\mathrm{rxe}}
\newcommand{\EZtxd}{E_\mathrm{tx0}}
\newcommand{\EBtxd}{e_\mathrm{txB}}
\newcommand{\EZrxd}{E_\mathrm{rx0}}
\newcommand{\EBrxd}{e_\mathrm{rxB}}
\newcommand{\Eil}  {E_\mathrm{lis}}
\newcommand{\LB}{L_\mathrm{phy}}
\newcommand{\Lsleep}{L_\mathrm{slp}}
\newcommand{\Lxsleep}{L_\mathrm{xslp}}

\newcommand{\Nempty}{n_\mathrm{emp}}

\usepackage{lipsum}
\usepackage[pscoord]{eso-pic}
\newcommand{\placetextbox}[3]{
  \setbox0=\hbox{#3}
  \AddToShipoutPictureFG*{
    \put(\LenToUnit{#1\paperwidth},\LenToUnit{#2\paperheight}){\vtop{{\null}\makebox[0pt][c]{#3}}}%
  }%
}%

\title{Enabling Listening Suspension in the Time Slotted Channel Hopping Protocol
\thanks{\noindent\hspace{-\parindent}978-1-6654-2478-3/21/\$31.00 
\textcopyright{}2021 IEEE}
}

\author{Gianluca Cena, Stefano Scanzio, and Adriano Valenzano\\
National Research Council of Italy (CNR--IEIIT), Corso Duca degli Abruzzi 24, I-10129 Torino, Italy\\
Email: \{gianluca.cena, stefano.scanzio, adriano.valenzano\}@ieiit.cnr.it\\
}

\begin{document}
\placetextbox{0.5}{1}{This is the author's version of an article that has been published in this journal.}
\placetextbox{0.5}{0.985}{Changes were made to this version by the publisher prior to publication.}
\placetextbox{0.5}{0.97}{The final version of record is available at \href{https://doi.org/10.1109/WFCS46889.2021.9483595}{https://doi.org/10.1109/WFCS46889.2021.9483595}}%
\placetextbox{0.5}{0.05}{Copyright (c) 2021 IEEE. Personal use is permitted.}
\placetextbox{0.5}{0.035}{For any other purposes, permission must be obtained from the IEEE by emailing pubs-permissions@ieee.org.}%

\maketitle
\thispagestyle{empty}
\pagestyle{empty}

\begin{abstract}
Time slotted channel hopping provides reliable and deterministic communication 
in IEEE 802.15.4 mesh networks.
Although slotted access is able to lower energy consumption drastically by reducing the 
duty cycle of the radio module, it usually leads to significant idle listening experienced
by receivers,
which makes it a sub-optimal solution when ultra low-power wireless is sought for.

In this paper a listening suspension mechanism is described,
which operates at the MAC layer and is part of a more general approach aimed at 
cutting down energy consumption by proactively reducing idle listening.
Links can be temporarily disabled, that convey 
slow-rate data streams whose characteristics, e.g., the generation period, 
are either known in advance to some extent or can be inferred by traffic inspection.
\end{abstract}

\section{Introduction}
\label{sec:introduction}
Time slotted channel hopping (TSCH) \cite{7451274, 9187609} 
is an enhanced access mechanism for IEEE 802.15.4 \cite{IEEE-802.15.4-2020} that,
thanks to scheduled access to the transmission medium, offers high reliability, determinism,
and, due to the reduced duty cycles, low power consumption \cite{2020-Sensors, 2013-Allerton}.
This enables the  connection of  battery-powered devices \cite{7812629}
in application scenarios like process and factory automation,
e.g., to retrofit industrial plants in order to include new features and functions \cite{2019-IA-fault_diagnosis}.
The medium access control (MAC) mechanism of TSCH relies on a periodically recurring slotframe,
to which all nodes must be precisely synchronized.
However,
an asynchronous transmission service is offered to the users of the data-link layer,
which is compatible with the Internet of Things (IoT) and Industrial IoT (IIoT) paradigms,
where the Internet Protocol (IP) is exploited at the network layer.
For this reason, some interesting IoT solutions have appeared recently, 
such as 6TiSCH \cite{2020-IETF-6TiSCH, 2014-BOOK-6TISCH-TCSH, 2020-6TiISCH-Survey},
which rely on TSCH  for data transmission.

In most wireless sensor networks (WSN) sensing is performed  periodically by motes.
Packet transmission in TSCH occurs cyclically too, 
but the sampling periods selected by applications are generally uncorrelated to the slotframe duration.
Therefore, network configuration and setup of applications can be decoupled,
and the design and deployment of distributed functions in heterogeneous systems (e.g., remote diagnostics and proactive maintenance) made easier.
This means that two distinct kinds of periodicity can be found in TSCH-based solutions like 6TiSCH. 
At the MAC level, the duration $\Tsf$ of the slotframe 
(in terms of number of slots) is selected network-wide.
At the application level, instead, a number of different sampling periods, 
denoted $T_{\mathrm{c},i}$, may be defined, each one depending on specific physical dynamics.
For example, the temperature of a liquid in a tank may vary more slowly than a flow through a pipe.  
Clearly, $\Tsf$ must be chosen so that it is smaller than the shortest period for variables in the network, $\Tsf \leq \min_i \{ T_{\mathrm{c},i} \}$. 

This is not the only constraint $\Tsf$ must satisfy.
Assuming that a single transmission opportunity (i.e., cell) is reserved in every slotframe for each pair of directly communicating nodes, to prevent network congestion
the overall traffic on any link must never exceed the slotframe repetition rate,
not even when interference and disturbance cause frame retransmissions.
Network parameters should be set so as to take into account the probability 
that a transmission attempt fails.
In practice, a safety margin has to be provided to avoid unwanted queuing phenomena in motes.
In the current version of OpenWSN for OpenMote B, the retry limit $R$ is set to $15$,
the default slot duration $\Tslot$ to $\unit[20]{ms}$,
and the slotframe size to $101$ slots (i.e., $\Tsf=\unit[2.02]{s}$).
Thus, sampling rates as low as half a minute can be quietly selected, 
network- and link-capacity wise.
 
In many applications, like condition-based monitoring involved in predictive maintenance,
sampling periods can be much longer than the slotframe.
If so, a non-negligible amount of energy is wasted in TSCH,
which can shorten the operating time significantly when motes are powered on batteries.
TSCH does not introduce any waste of energy on the transmitting side of a link.
In fact, if there is no packet queued for transmission when a slot becomes available, 
no attempt is performed and the cell is simply left unused.
By contrast, the receiving side of the link must be enabled in every cell it is associated,
since it must be ready to get each frame potentially transmitted.
As a consequence the receiver mote must be up and listening for some portion of the slot, 
in order to establish whether or not a transmission is being performed by some nearby peer.
Unfortunately, in many real situations listening to the channel when a cell is unused is comparable to a frame reception from the power consumption viewpoint.
This phenomenon is known as \textit{idle listening}, and is the cause of energy wastes.

A simple solution, when all sampling periods are much larger than the slotframe duration,
is the reconfiguration of the MAC protocol by increasing the number of slots in the slotframe 
($\Tslot$ depends on the physical layer and can hardly be changed).
This introduces a drawback affecting cell $\left\langle 0,0 \right\rangle$, 
that is the cell placed at slot and channel offsets $0$ in the TSCH matrix,
which is used by 6TiSCH for neighbor discovery and RPL topology construction activities.
In fact, network reconfiguration (e.g., when some motes are moved or switched on/off, obstacles are placed between them, nearby sources of interference and noise appear/disappear) becomes slower as $\Tsf$ increases.
In many cases this limitation is not severe since, from a practical viewpoint, 
having network reconfiguration times much shorter than the sampling period 
makes little sense. However, when sampling periods are very different 
(e.g., when they range from few minutes to several hours) enlarging $\Tsf$ may not suffice. 
This is because $\Tsf$ has to be selected starting with the highest dynamic packet flow,
which makes this approach useful only in part for the slower streams. 
Moreover, excessive increases of $\Tsf$ reduce the ability of nodes to keep  their time sources synchronized, which is a basic requisite for correct time slotting operation.

A more effective solution to this problem, which enables ultra-low power communication,
consists of suspending the receiving mote when no frame is expected to arrive, so that no energy is wasted because of idle listening.
Such a kind of technique, named Proactive Reduction of Idle Listening (PRIL),
was introduced in \cite{10.1007/978-3-030-61746-2_11}.
It is worth noting that acting this way is not trivial, as the involved motes must know  in advance when transmissions occur,
and this is in contrast with the IoT paradigm where packet transmission requests are driven by applications.
A practical method is to explicitly drive the \textit{listening suspension} (LS) of the link receiver by the transmitter,
through the inclusion of suitable \textit{sleep commands} in the frames being sent.
This approach, which requires slight changes to the TSCH MAC that do not impair backward compatibility,
was envisaged in \cite{10.1007/978-3-030-61746-2_11} by exploiting suitable information elements, embedded in IEEE 802.15.4 frames, to put the receiver to sleep.
In this way the transmitter holds all the knowledge about data exchanges,
while the receiver simply obeys to sleep commands.
Doing so prevents inconsistencies between link sides and tangibly improves LS reliability.

Mechanisms employed at the data-link layer to disable listening in selected cells should be clearly separated from the policies defined at higher levels to reduce the amount of idle listening.
In this paper we focus on the former issue, by defining commands to be included 
in the MAC, as well as a number of relevant strategies. Our goal is to provide 
some hints and bounds on the benefits that LS can offer in realistic scenarios.
Characterization of traffic patterns over links produced by applications and devices,
and the analysis of optimal LS strategies are left for future work.

The paper is structured as follows:
Section~\ref{sec:RIL} introduces the problem of reducing idle listening, 
whereas in Section~\ref{sec:LS} sleep commands and some related strategies are described.
Section~\ref{sec:PA} focuses on power saving achievable with different strategies,
while some numerical results are reported in Section~\ref{sec:PC}.
Finally, Section~\ref{sec:CONC} concludes the paper.

\section{Reducing Idle Listening} \label{sec:RIL}
Information about packet transfer timing can be obtained in two different ways:
it is either provided \textit{explicitly} by the user,
who instructs the data-link layer about its traffic flow,
or the data-link layer tries to \textit{implicitly} infer the pattern of exchanges by continuously analyzing the traffic flowing through the mote.
Both approaches are made harder as intermediate nodes in a mesh WSN forward packets produced by ascendants and descendants.
In general, combining implicit and explicit data produces best results.
Information about traffic can be obtained from different layers.
For example, the network layer permits to identify streams coming from or directed to specific endpoints along a route.
This is useful to decompose the traffic flowing through intermediate relays into separate and simpler contributions.
In a similar way, application processes know the characteristics of the traffic they produce and consume (e.g., periodicity, minimum inter-arrival times, deadlines).

\begin{figure}[b]	
    \vspace{-0.4cm}
	\centering
	\includegraphics[width=\columnwidth]{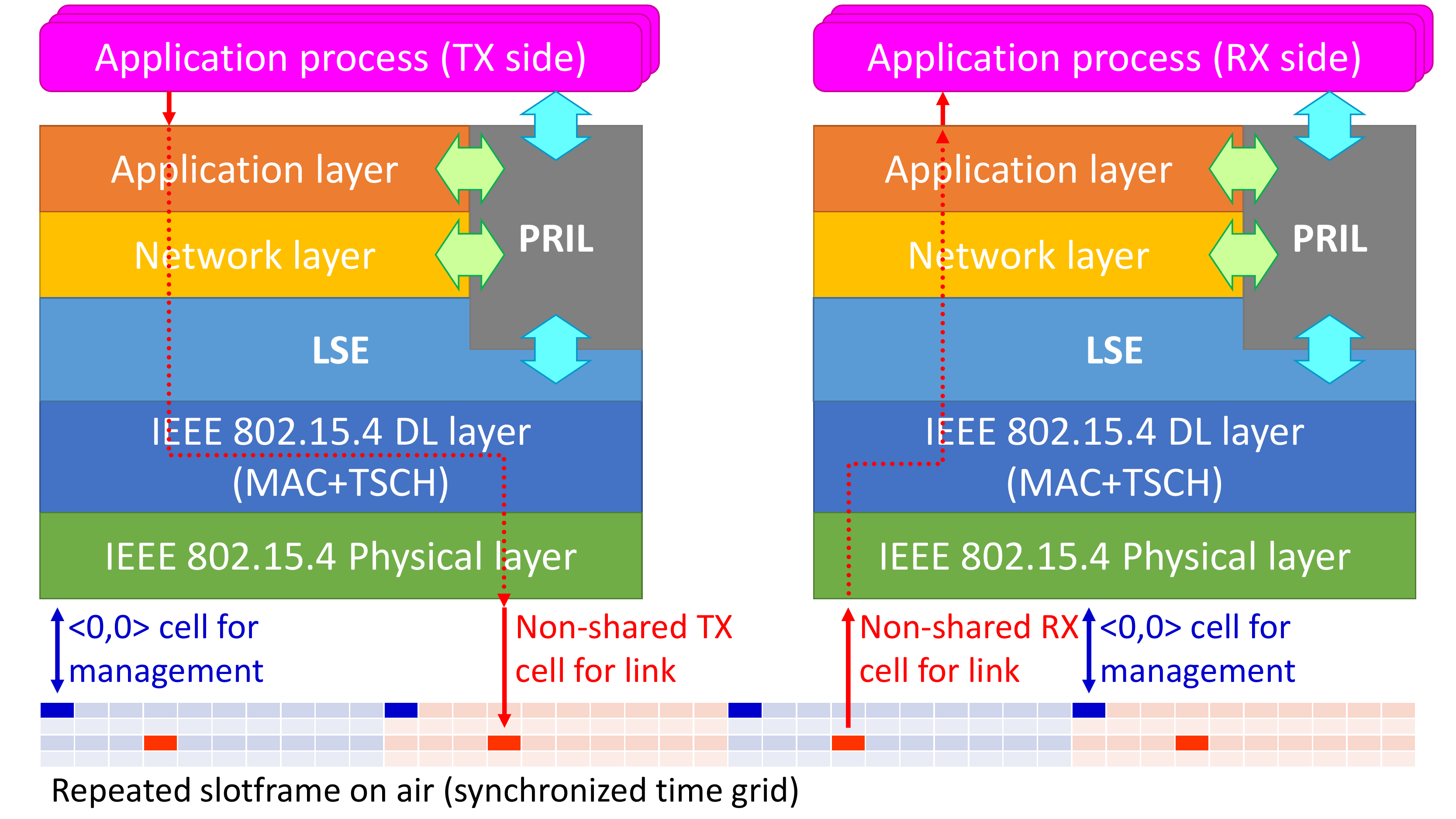}
	\caption{Block diagram of PRIL techniques (interacting with the LSE).}
	\label{fig:pril}
\end{figure}

Effective implementations consist of two separate and cooperating elements: 
PRIL and Link Suspension Entity (LSE).
PRIL concerns network management and mechanisms to establish dynamically at runtime when and how long a link can be safely suspended, in such a way that application constraints (e.g., latency, throughput) are not violated for the packet flow.
PRIL can include:
\begin{itemize}
\item 
\textit{Link traffic model} to describe traffic patterns conveyed through the link. 
For example, a single periodic stream is completely characterized by its period, while
a superposition of periodic streams is identified by the set of related periods and phase offsets. A sporadic stream can be defined by means of its average and minimum interarrival times.
\item 
\textit{Interfaces} to protocol layers.
The MAC layer interface is of utmost importance and enables interactions with the LSE.
Another interface is offered to application processes,
and is meant to support explicit LS configuration.
For instance, it allows the selection of a specific traffic model (periodic, sporadic, etc.),
the specification of its parameters (e.g., the sampling period of sensors),
and the definition of additional constraints known by applications only
(e.g., relative deadlines for sporadic messages).
\textit{Dissectors} are also envisaged for implicit traffic characterization.
They  enable analyses of information in protocol headers, 
which are encapsulated in frames sent over the link
(e.g., IP source and destination addresses, UDP source and destination ports).
\item 
\textit{Algorithms (intelligence}), possibly based on machine learning,
to (semi-)autonomously detect traffic patterns from data mentioned above 
(i.e., finding a traffic model and its parameter values fitting the link)
and optimally drive the LS mechanism in the MAC.
\end{itemize}

Instead, LSE is part of the MAC and consists of:
\begin{itemize}
\item
LS-related \textit{commands}, to be embedded in exchanged frames. 
They include requests issued by the transmitter  to temporarily disable listening on the receiver and other elements involved in specific PRIL techniques.
\item
Protocol finite state machines and local state variables implementing the 
LS mechanism and supporting a set of pre-defined LS \textit{strategies}.
\item
Data-link layer \textit{primitives} to enable interactions between PRIL and LSE.
They should be simple, flexible, and powerful enough to cope with different strategies.
\end{itemize}

\section{Listening Suspension Mechanism}
\label{sec:LS}
The LS mechanism for TSCH belongs to the MAC layer.
New primitives have to be defined in the MAC sublayer management entity (MLME) to interact with the LSE.
For example, a generic MLME-LS-SET can be designed to enable/disable LS on a given link,
to select a specific strategy,
and to specify the strategy operating parameters.
A detailed specification of new primitives cannot abstract from a stable definition of traffic models and PRIL interfaces, which are not available yet.
In addition, the behavior of data transmission primitives of the MAC common part sublayer (MCPS), 
such as MCPS-DATA, must also be modified. 
On the one hand, in fact, procedures involved in listening suspension have to be executed when needed (e.g., triggered by either transmission requests or local timers and counters)
while, on the other hand, traffic needs to be analyzed by PRIL starting from transmission requests
(e.g., by setting function hooks to inspect frames on the link), to infer its characteristics.

In the current proposal sleep commands are defined as a part of the LSE
and sent by the transmitter to the link receiver in order to force the latter to temporarily disable listening.
They are conveyed in data frames by using information elements (IE),
that are special fields that can be optionally included in IEEE 802.15.4 frames to carry ancillary information.
IEs extend the basic protocol but preserve backward compatibility with existing devices and networks.
In defining a new IE for sleep commands, denoted \textit{sleep IE}, 
a streamlined encoding must be envisaged so as not to undermine energy saving these mechanisms are intended to achieve.
In fact, the inclusion of IEs in frames increases the number of bits sent on air and, consequently, the power consumption on both the transmitting and receiving sides.

Every sleep command only concerns the link where the including frame was received.
This is because any intermediate mote acting as a relay can have multiple links connected to its children and/or parent neighbors, and suspension must be managed independently for each one of them.
It is worth noting that the LS mechanism works on both the link sides, 
by either putting them on hold or restoring the conventional TSCH behavior.
This prevents the sender from performing transmission attempts when  listening is temporarily disabled on the receiver.
Of course, suitable variables have to be defined on the transmitting and receiving sides of every link to keep track of its LS state (they do not refer to the motes as a whole).

In the following subsections some possible options are briefly described for LS commands 
and for general-purpose LS strategies relying on them.

\subsection{Basic Sleep Command}

\begin{figure}[b]
	\centering
	\includegraphics[width=\columnwidth]{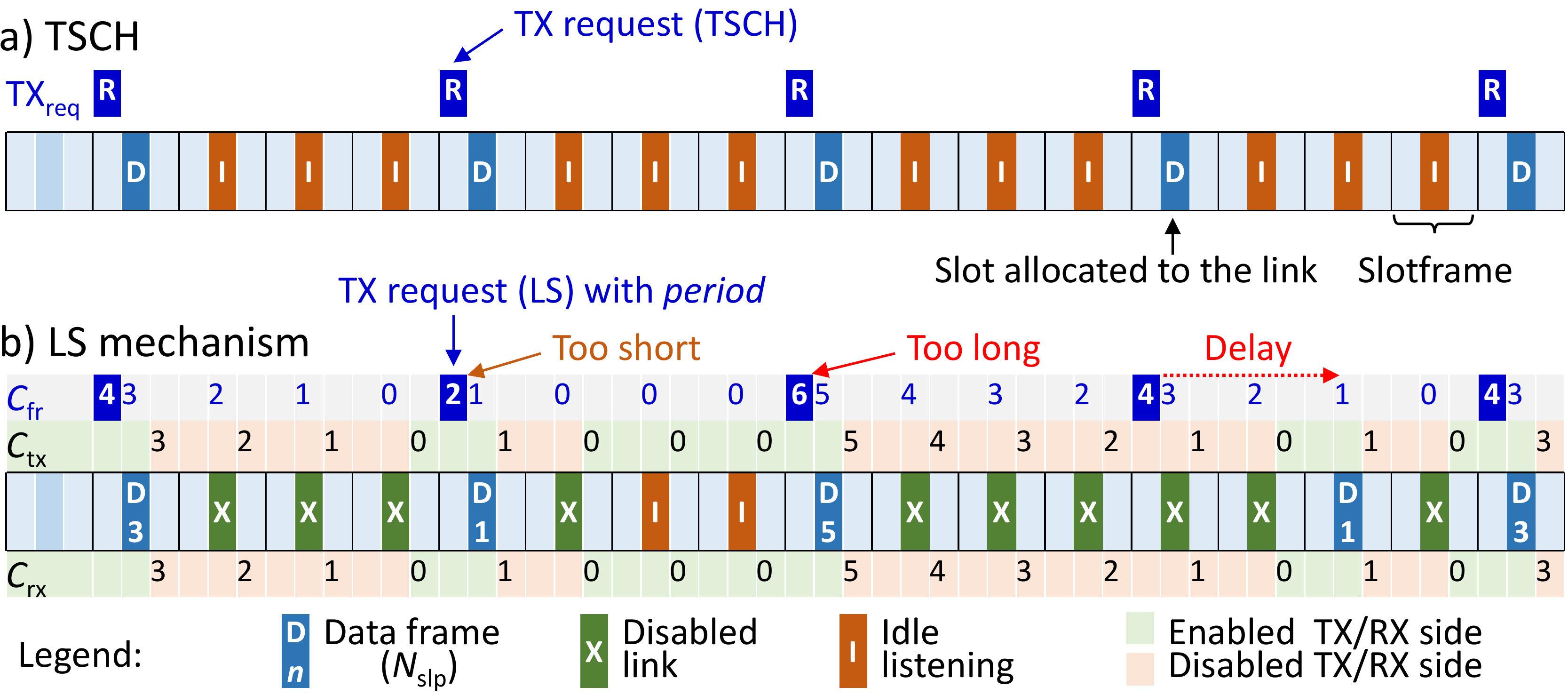}
	\caption{Conventional TSCH vs. LS mechanism.}
	\label{fig:slp1}
\end{figure}

The simplest way to instruct the receiver about the duration of the listening suspension 
is by specifying a \textit{sleep time} in the sleep IE.
This time is encoded as a small integer, denoted $\Nsleep$,
representing the number of slotframes where listening has to be suspended 
(note that the final decision about the suspension is left to the receiver,
which ensures backward compatibility).
Upon reception, the link is disabled during the following $\Nsleep$ slotframes, 
then it is enabled again, restoring the conventional TSCH operation.
Fig.~\ref{fig:slp1} shows that the time interval before the next usable cell of the link is $(\Nsleep+1) \Tsf$ wide.
In the figure, the number of cells that, in every slotframe, precede and follow 
the cell allocated to the link is fixed but arbitrary.

The special value $\Nsleep=0$ informs the receiver that it must be listening in the next slotframe, and it is roughly equivalent to not including any sleep command.
However, while the absence of sleep commands has clearly no effects,
setting $\Nsleep=0$ can be used to trigger some specific actions.
For example, when multiple cells are reserved for the same link in the slotframe, 
this can enforce an intra-slotframe LS behavior by skipping all the remaining 
cells in the current slotframe.
This feature, which is useful when overprovisioning is exploited for high-capacity links 
\cite{9114449}, is not considered in the remaining part of this paper.
Values $\Nsleep \geq 1$ are instead meant to enforce inter-slotframe listening suspension.

Another point concerns the maximum extension of the sleep time interval.
In our opinion $6$ bits are enough to encode $\Nsleep$ for most real situations, leading to 
link suspension periods up to about two minutes.
Allocating more bits to $\Nsleep$ does not bring noticeable advantages from the point of view of power consumption.
In fact, the shared cell $\left\langle 0,0 \right\rangle $ remains always active in every slotframe, 
so that longer sleeping periods for other cells become irrelevant for energy saving 
(unless the mote is involved in several links).
In addition, a receiving cell disabled by a sleep command cannot be (easily) re-enabled before the re-activation time is reached. 
Responsiveness to alarms and operations taking place on demand, such as firmware upload or configuration over the air, worsens and may become unacceptable when too long sleep intervals are enforced.

LS works properly when appropriate values are selected for $\Nsleep$ at any time.
For example, if the link bears a single periodic packet stream,
$\Nsleep$ should be chosen to match the packet generation period.
As mentioned before, characterization of the traffic model and its parameters is up to PRIL,
since the MAC layer has to be kept as simple as possible.
Suitable primitives can drive intelligence in collecting information: for instance, some parameters (e.g., deadlines) can be provided by upper layers while others (e.g., periods) can be inferred by analyzing the traffic over the link.

\subsection{Periodic LS Strategy}

\begin{figure}[b]
	\centering
	\includegraphics[width=\columnwidth]{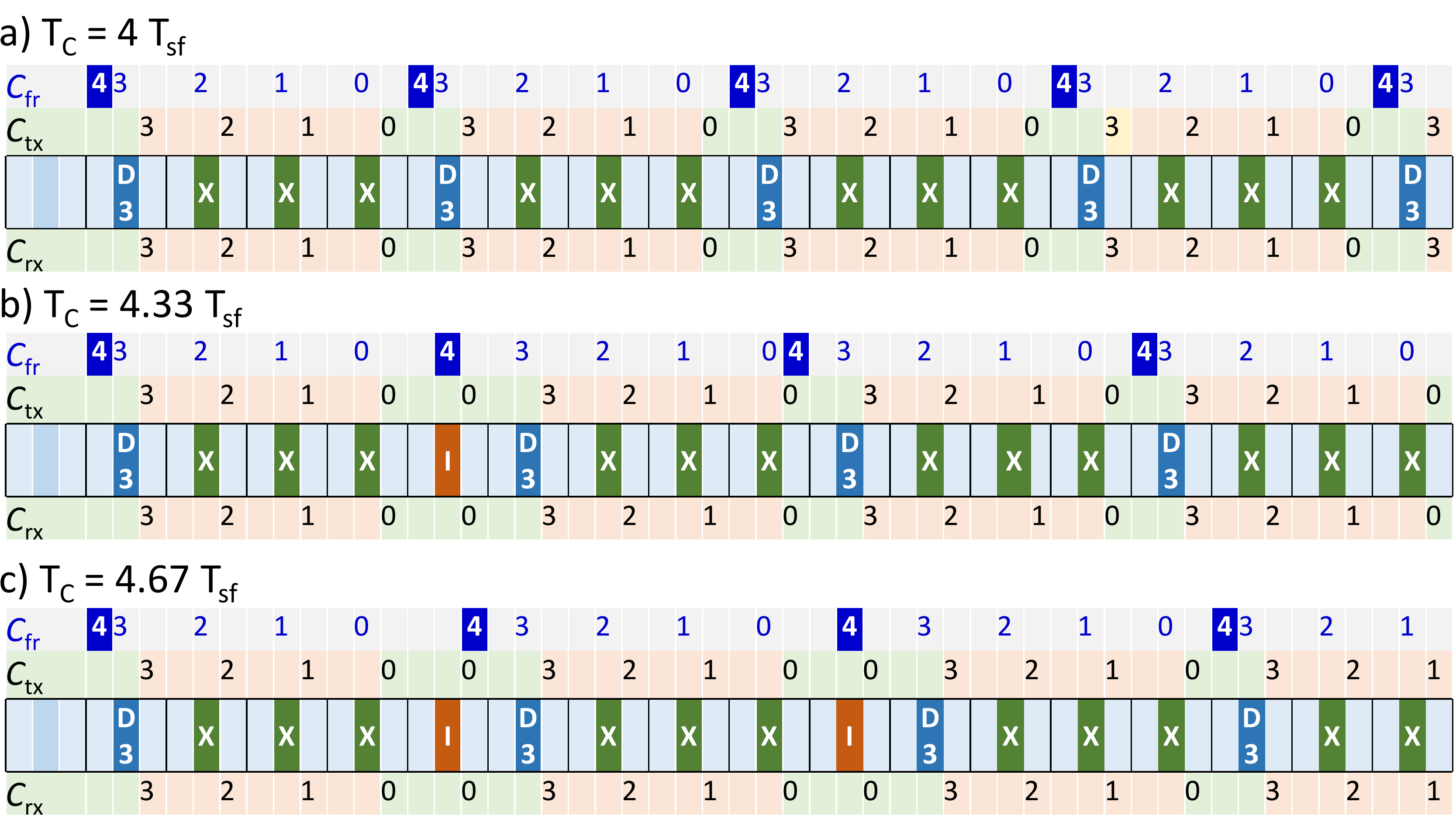}
	\caption{Periodic LS strategy (by varying the actual period).}
	\label{fig:slp2}
\end{figure}

Let us first consider a simple LS strategy fitting the needs of periodic traffic completely defined by its period $\Tc$.
Under the assumption that only one cell is reserved per link, network stability requires that $\Tc > \Tsf$  (strict order applies, as transmission errors are unavoidable).
Whenever a new frame transmission request is issued for the link 
(e.g., through the MCPS-DATA.request primitive),
the transmitter starts a local \textit{frame sleep} counter $\Ctrq$ for the frame. 
The counter is initialized to $\Ctrq = \lfloor \tauc \rfloor \geq 1$,
where $\tauc = \Tc / \Tsf$ is the normalized transmission period. It is worth noting that 
rounding down allows to re-enable the link and, in particular, the receiver listening 
before a new packet becomes available on the transmitter, thus reducing both the access time and the average number of queued packets.
Of course, if  $\Tc$ is not a multiple of $\Tsf$, 
idle listening may occasionally take place when the receiver is re-enabled,
as depicted in Fig.~\ref{fig:slp2}.

When the frame is sent, the value of $\Nsleep$ in the sleep IE is set to $\Ctrq$, moreover
the sleep command is included in the frame only if $\Nsleep>0$.
Reception of a sleep command causes the receiver to set its own \textit{reception sleep} counter $\Crls$ to $\Nsleep$.
Upon ack frame reception, the $\Ctrq$ value is also loaded into the \textit{transmission sleep} counter $\Ctls$.
At the very beginning of every slot allocated to the link $\Ctrq$ is decreased by one, provided it is greater than $0$ (counters contain non-negative values).
The same is true, at the end of the slot, for $\Ctls$ and $\Crls$, assuming that they were not set in the same slot.
Each link side is assumed to be either enabled when the corresponding counter 
($\Ctls$ for the transmitter and $\Crls$ for the receiver) is equal to $0$,  or disabled otherwise.
Since slotframes are kept synchronized all over the network by TSCH,
$\Ctls$ and $\Crls$ are updated coherently on both sides of the link and their values are always the same.
Should synchronization between motes be lost, all sleep counters are reset to $0$, 
hence deactivating any ongoing listening suspension (and restoring conventional TSCH operation).

Frame delivery for a given transmission request may be occasionally delayed as shown, for example, to the fourth request in Fig.~\ref{fig:slp1}.b,
because of a prior command with an excessively large sleeping period.
This event is dealt with by the LS mechanism in this way:
if, for any reasons, the frame is not sent at the first opportunity following the transmission request and it has to wait in the local queue, $\Ctrq$ is correctly decreased by the LS algorithm.
Consequently, if the sleep command eventually reaches the receiver, the re-enable instant is not affected. 
This also copes with the relevant case of failed transmission attempts (detected by expiration of the ack timeout), where $\Ctrq$ is decreased by one on each retry.
In this way, transmission errors affect the current packet in the same way as TSCH,
whereas the next packet does not suffer from additional delays with respect to the case when no errors occur.

To help recovering from queuing phenomena, 
which may take place for instance when long bursts of errors are experienced,
the sleep command is skipped if more than one frame concerning the same link is pending in the transmission queue, thus preventing the link capacity from being throttled in these conditions
(which could lead to buffer overruns). 

This strategy can also be used for quasi-periodic traffic,
where packet generation takes place cyclically at a nominal rate but release times are subject to non-negligible jitters.
If the time interval between two consecutive packet transmission requests exceeds $\Tc$,
receiver idle listening occurs.
Conversely, when the interval is shorter than $\Tc$,
the early packet experiences some delay since the transmission can start only when the link is re-enabled (i.e., $\Ctls=0$ and $\Crls=0$).

\subsection{Responsiveness}
Responsiveness in TSCH is limited by its MAC mechanism, which is based on a continuously repeating slotframe \cite{2016-WCN}.
As pointed out in \cite{AHN2020}, in absence of errors due to frame corruption on air,
one-way transmission latency consists of two elements, namely access delay and transmission time.
Access delay is the time the transmitter spends waiting for the slot assigned to the link, 
and is the largest contribution.
Under the assumption that, at any time, at most one frame is queued for transmission for the link, it can be modeled as a random variable uniformly distributed between $0$ and $\Tsf$.
The transmission time includes the offset at the beginning of the slot (\textit{macTsTxOffset})
as well as the time taken for transmitting the data frame on air.
Since it is always strictly shorter than $\Tslot$ we will neglect it in the following.
In this way
the \textit{worst-case transmission latency} $\Twc$  is approximately the same as the maximum access delay when no errors occur.

The access delay may grow sensibly when sleep commands are exploited.
With respect to conventional TSCH the period between usable cells for a link (and access delays) increases from $\Tsf$ to $(\Nsleep+1)\Tsf$,
while energy spent by the receiver for listening to the channel decreases by about the same factor.

With strictly periodic patterns of packets and a proper selection of $\Nsleep$ (ensuring that the link is re-enabled just before a new packet becomes available for transmission) 
delays are roughly the same as TSCH.
However, packets generated sporadically can experience access delays up to $(\Nsleep +1) \Tsf$ when they become ready just after the link suspension.
To improve link responsiveness for sporadic packets, 
$\Nsleep$ can be set based on their \textit{relative deadline} $\Tmin$.
In particular, to ensure that $\Twc \leq \Tmin$, again in the optimistic case of an ideal channel without errors, $\Nsleep$ has to be set so that $\Nsleep \leq \taud-1$, 
where $\taud = \Tmin/\Tsf$ is the normalized deadline.
For example, with the network parameters we considered,  $\Nsleep$ must not exceed $13$ if latency must be shorter than $\unit[30]{s}$.
If the period is sensibly larger than the deadline ($\Tc \gg \Tmin$),
the link is re-enabled too early most times and the LS mechanism behaves less effectively.

\subsection{Slow Periodic LS Strategy}
In slow periodic streams the period may exceed the maximum value allowed for the sleep command
($64 \Tsf$ for $\Nsleep$ 6-bit encoding).
In such cases the LSE takes care of issuing sleep commands automatically to repeatedly restart the sleeping period in the receiver and make the suspension last the desired amount of time.
$\Nsleep$ is equal to its maximum value ($63$) in all the sleep commands in the sequence except for the last one, 
which is set to $( \lfloor \tauc \rfloor -1) \bmod 64$ and re-enables the link at the intended time.
Overall, the sequence includes $\lceil \tauc / 64 \rceil$ frames.
For example, if $\Tc$ is equal to $10$ minutes (i.e., $\Nsleep=296$),
five frames are sent per packet, spaced by $64 \Tsf$:
the first one is the data frame, which includes the packet and a sleep command with $\Nsleep=63$,
followed by three empty frames (with no payload) where $\Nsleep=63$,
and one final (empty) frame where $\Nsleep=39$.
Should a new transmission request be issued before the above procedure is completed,
the course of action is interrupted by the transmitter and the new data frame is sent as soon as the link is re-enabled.

The impact of this approach on power consumption is relatively low 
and is roughly equivalent to the energy needed for sending and receiving one frame every $64$ slotframes (about two minutes). 
Most of these frames contain no actual data and only deliver sleep commands to the receiver.
To further reduce power consumption a specific short format can be envisaged for 
\textit{empty sleep frames} that only carry the sleep IE. A new MAC sleep command frame can be defined to this purpose but other solutions are possible as well.
Note that from the energy viewpoint this contribution is rather negligible when compared to cell 
$\left\langle 0,0 \right\rangle$, which is active in every slotframe,
nonetheless it improves network responsiveness by imposing a reasonable upper bound 
$\Twc=64\Tsf$ on transmission latency.

\subsection{Extended Sleep Command}

\begin{figure}[b]
    \centering
    \includegraphics[width=\columnwidth]{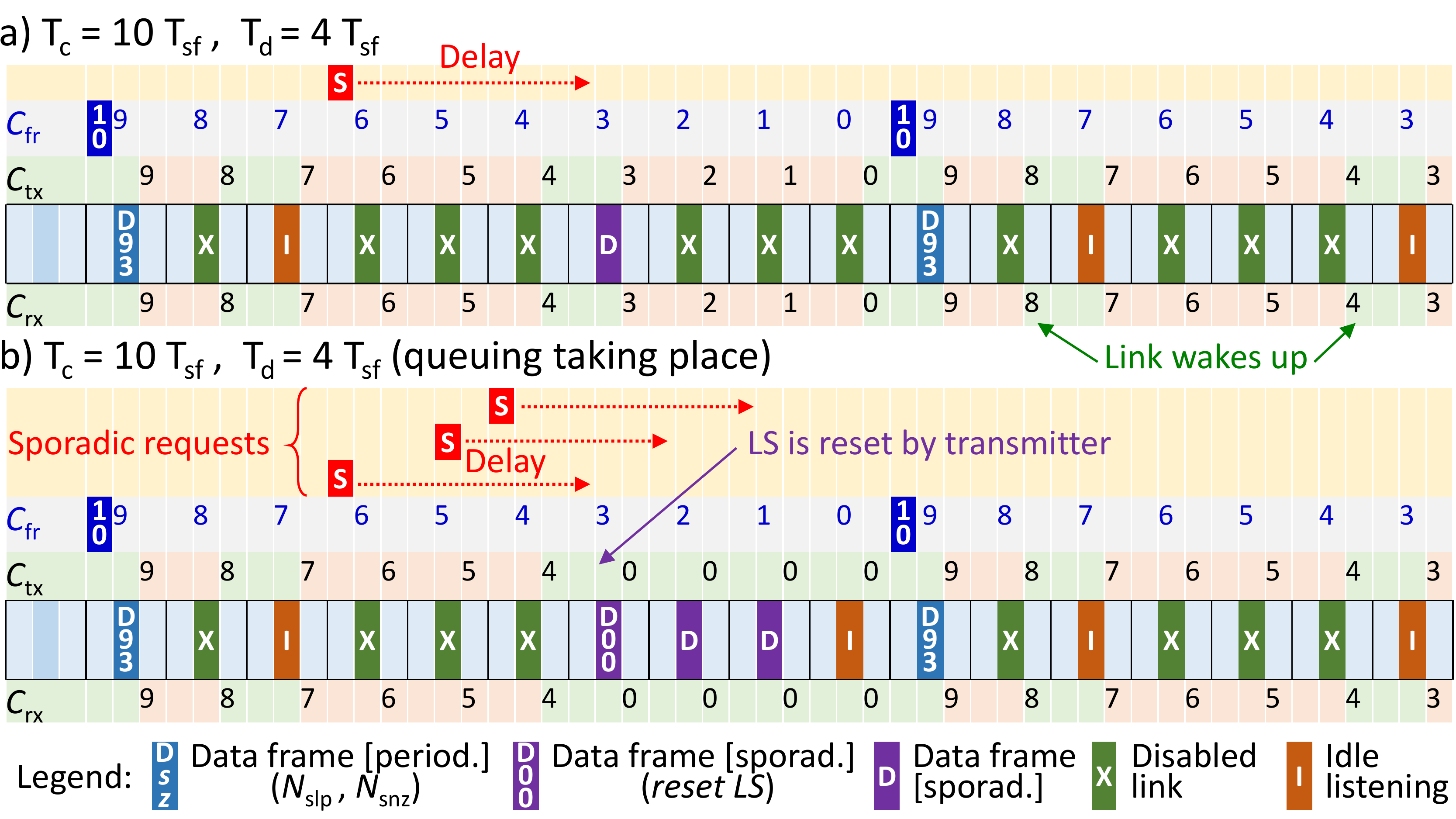}
    \caption{Extended periodic LS strategy (with and without queuing).}
    \label{fig:slp3}
\end{figure}

Energy consumption can be reduced without impairing responsiveness too much, by means of
an extended version of the sleep command we call \textit{xsleep}.
The relevant IE includes two parameters: $\Nsleep$ and $\Nsnooze$.
$\Nsleep$ is the number of slotframes the link must be kept disabled. It specifies the \textit{nominal} duration of the  sleeping period and is managed the same way as in basic sleep commands. 
However, the range of allowed values for the sleep time can be increased noticeably in this case
without latency is consequently worsened.  
For instance, $\Nsleep$ can be encoded with $12$ bits and the sleeping period can last up to $4096\Tsf$ (more than two hours). Selection of larger field sizes to enable longer sleeping periods is clearly possible.

The \textit{snooze} time, denoted $\Nsnooze$, specifies how often the link must be temporarily re-enabled (\textit{awakened}) during the nominal sleeping period,
in order to flush packets sporadically generated in the meanwhile and left pending in the transmission buffer.
In this case, on correct reception of the \textit{xsleep} command an iterative procedure is started, where the link is disabled for $\Nsnooze$ consecutive slotframes before being awakened for one slotframe.
This behavior is automatically repeated by the LSE until the end of the nominal suspension period (as specified by $\Nsleep$) is reached.
Clearly, $\Nsnooze$ must be strictly lower than $\Nsleep$.
$\Nsnooze$ encoded with $6$ bits (e.g., $\Nsnooze \leq 63$) provides upper bounds for the access time similar to the basic sleep command.
In practice, $3$ bytes are enough to encode the \textit{xsleep} command into the content of a sleep IE.

Identification of slotframes where the link must be awakened does not consider the instant when the sleep command reaches the receiver, but proceeds from the end of the sleeping period moving backwards.
In other words, frames can be sent only when $\Ctls \bmod (\Nsnooze+1) = 0$.
In this way delays in setting $\Crls$ caused, for instance, by transmission errors do not affect the wake-up synchronization on the two link sides.
For example, if $\Nsleep = 58$ ($\Tc \simeq \unit[120]{s}$) and $\Nsnooze=13$ ($\Tmin \simeq \unit[30]{s}$),
the receiver temporarily wakes up at the $3$-rd, $17$-th, $31$-st, and $45$-th slotframes after the command is issued,
and it is re-enabled at the $59$-th slotframe.

Fig.~\ref{fig:slp3} assumes $\Nsleep = 9$ and $\Nsnooze=3$,
and shows that awakening occurs when counters equal $8$ and $4$.
If needed, the \textit{xsleep} command can be aborted at those time instants, 
for instance when queuing occurs as in Fig.~\ref{fig:slp3}b.
If, for whatever reason, 
a number of sporadic packets become pending in the transmission buffer (e.g., as a consequence of prolonged interference on air),
the transmitting LSE sends a new sleep command to reset the behavior of the receiver.
A special \textit{xsleep} command with $\Nsleep=0$ and $\Nsnooze=0$ can be defined to this purpose, 
which reverts the behavior to conventional TSCH. 
Then, queued packets can be drained at the maximum link speed and LS operation resumed when the queue is empty.

LS strategies that rely on \textit{xsleep} trade energy for responsiveness.
In fact, the worst-case access delay (and $\Twc$ as well) is equal to the \textit{snoozing period} $(\Nsnooze+1)\Tsf$.
Selecting a low value for $\Nsnooze$ reduces $\Twc$, but also increases power consumption on the receiver side consequently.
Because of the large values allowed for $\Nsleep$ (latency now only depends on $\Nsnooze$)
the time between transmissions of \textit{xsleep} commands can be as large as about two hours, 
which makes energy consumption completely negligible on the transmitter side.

\subsection{Extended Periodic LS Strategy}
This strategy is suitable for a traffic pattern consisting of the superposition of 
a quasi-periodic stream with (average) period $\Tc$ and sporadic streams with relative deadline $\Tmin$.
For simplicity we assume that sporadic packets, which should be served promptly,
are issued in consequence of rare events.
In this case, \textit{xsleep} commands are conveyed inside data frames carrying periodic packets, 
and $\Nsnooze$ and $\Nsleep$ are selected so that 
$\Nsnooze = \lfloor \taud \rfloor - 1$ and $\Nsleep = \lfloor \tauc \rfloor - 1$.

Such a strategy closely resembles periodic LS,
but the receiver wakes up regularly during the sleeping period to ensure bounded transmission latency to sporadic packets.
Since \textit{xsleep} enables large sleeping times, high efficiency can be reached.
Moreover, should the inferred period be updated by PRIL, 
changes can be enforced by sending a new \textit{xsleep} command as soon as the receiver awakens.

\section{Performance Analysis}
\label{sec:PA}
Energy consumed on the transmitting and receiving link sides can be evaluated for conventional TSCH and LS strategies.
In the following we assume that no transmission error occurs, to simplify the analysis and focus on the intrinsic energy saving capability of the proposed solutions.

\subsection{Power Consumption Model}
Let $\Etxd$ and $\Erxd$ be the amounts of energy spent for performing a single data frame transmission and reception attempt, respectively.
In the following, we model these quantities by assuming that they depend on the frame size linearly.
In particular, $\Etxd = \EZtxd + \EBtxd \cdot \LB$,
where $\LB$ is the overall number of bytes transmitted by the physical layer,
$\EBtxd$ is the energy required to send one byte, and $\EZtxd$ is a constant amount of energy additionally spent for every transmission attempt.
For instance, for OpenMote B boards running OpenWSN and communicating through 6TiSCH,  
$\EZtxd = \unit[7]{\mu J}$ and $\EBtxd = \unit[2]{\mu J/B}$ (see plots of Fig.~4 in \cite{9187609}), thus the largest fraction of power directly depends on the number of bytes transmitted on air. 

A very similar relation holds for reception.
From the same set of experimental data we obtain $\EZrxd = \unit[65]{\mu J}$ and $\EBrxd = \unit[1.3]{\mu J/B}$.
Unsurprisingly, the energy per byte is lower than for transmission.
However, a non negligible amount of energy is spent because listening must be enabled for some time preceding 
the instant when transmission is expected to begin.
$\EZrxd$ is a mean value, and fluctuations can be experienced depending on the alignment precision of the transmitter and receiver time grids at any given time.

Energy needed to send and receive the ack frame ($\unit[33]{B}$)
is $\Etxa = \unit[106]{\mu J}$ and $\Erxa = \unit[79]{\mu J}$, respectively.
This also includes operations to finalize a frame exchange besides the transmission/reception on air.
Finally, the energy spent every time idle listening occurs for the cell is $\Eil = \unit[138]{\mu J}$.
$\Eil$ is about twice $\EZrxd$, since the listening interval is centered around the expected start of frame transmission.

In absence of frame transmission and reception the mote drains about $\unit[31.4]{mW}$ (i.e., $\unit[628]{\mu J}$ in every slot).
This is due to the basic activity of the operating system, protocol stack and application processes,
and concerns all board components besides the microcontroller.
It is worth observing that such a high value is likely due to the fact that both the OpenWSN software and the OpenMote B hardware have not been optimized yet.
Hopefully, this constant consumption will be lowered dramatically when industry-grade devices based on this technology are designed and implemented.
For these reasons this contribution in not considered in the following analysis, which
focuses only on energy spent for communication.

\subsection{Conventional TSCH}
Let $\Lambdasf=1/\Tsf$ and $\Lambdac=1/\Tc$ be the slotframe repetition rate and the packet transmission rate, respectively.
The transmitter power consumption for a packet stream sent over a conventional TSCH link is
\begin{align}
	\Ptx = ( \Etxd + \Erxa ) \Lambdac.
\end{align}
When only reception (and acknowledgment) of frames is considered, for the receiver we have
\begin{align}
	\Prz = ( \Erxd + \Etxa ) \Lambdac.
\end{align}
$\Prz$ equals the power spent by the receiver when an oracle-based LS strategy is adopted.
Thanks to the oracle, listening is enabled only when the cell conveys a frame,
without the need to obtain any information from the sender by means of sleep commands embedded in IEs. 
As a consequence, $\Prz$ provides a lower bound to power consumption for a given stream.
The overall receiver consumption is given by
\begin{align}
	\Prx =  \Prz + \Eil [\Lambdasf - \Lambdac],
\end{align}
where the rightmost term refers to idle listening, which occurs when the cell allocated to the link is not used for data (hence, the difference between $\Lambdasf$ and $\Lambdac$).

In a TSCH/6TiSCH network operating correctly, additional frame exchanges are performed automatically in every slotframe.
This is the case of transmission/reception in the shared cell $\left\langle 0,0 \right\rangle$,
which is meant for network management activities. 
Despite their contribution to power consumption, they are not taken into account here because they are not correlated with data streams produced by applications and affected by the LS mechanism.
Roughly speaking, power consumption for these frame exchanges can be expressed as 
$P_0 = \bar{E}_{\left\langle 0,0 \right\rangle }  \Lambdasf$,
where $\bar{E}_{\left\langle 0,0 \right\rangle }$ is the average energy spent by the mote
to carry out one (non-confirmed) transmission/reception attempt in cell $\left\langle 0,0 \right\rangle$.
In practice, the value of $\bar{E}_{\left\langle 0,0 \right\rangle }$ does not differ from $\Etxd$, $\Erxd$, and $\Eil$ significantly.

\subsection{Periodic LS Strategy}
In periodic LS a basic sleep command is included in every data frame where $\Nsleep = \lfloor \tauc \rfloor -1$.
For example, if $\Tc=\unit[30]{s}$ then $\Nsleep=13$.
In the following we will assume that $\Tc$ can be inferred precisely by the intelligence of the PRIL module.
Obviously, this represents the best case for the LS mechanism, and enables the computation of an upper bound for improvements this approach can offer.

Power consumption on the transmitting side increases slightly because of the sleep command
\begin{align}
	\Ptx^\mathrm{P} 
	&= \Ptx + \Lsleep \EBtxd \Lambdac,
\end{align}
where $\Lsleep$ is the size in bytes of the encapsulating IE ($\Lsleep=\unit[3]{B}$ in our case).
For typical frame size (\unit[90]{B}) the increase is negligible ($\sim 3\%$).

Data frames that carry sleep commands suspend listening for $\Nsleep$ slotframes when they are received correctly.
Thus every frame prevents idle listening from occurring $\Nsleep = \lfloor \tauc \rfloor -1$ times. 

Hence
\begin{align} 
	\Prx^\mathrm{P} \nonumber
	&= \Prz + \Lsleep \EBrxd \Lambdac + \Eil \left[ 
	\Lambdasf - \left\lfloor \tauc \right\rfloor \Lambdac \right] \\
	&= \Prx + \Lsleep \EBrxd \Lambdac - \Eil \left[ \left\lfloor \tauc 
	\right\rfloor -1 \right] \Lambdac.
\end{align}
Again, the consumption increase due to sleep IEs is typically negligible.
Instead, the contribution due to idle listening (rightmost term in the first line of the equation) is considerably lower than TSCH, 
because the multiplicative factor for $\Eil$ is always smaller than $\Lambdac$,
and equal to $0$ if $\Tc$ is a multiple of $\Tsf$. 
In the latter case the difference in terms of power consumption between the oracle and the periodic LS strategy is constant and equal, overall, to $\Lsleep(\EBtxd+\EBrxd)\Lambdac$,
which, for $\Lsleep=\unit[3]{B}$ and $\Tc=\unit[30.3]{s}$, means $\unit[0.33]{\mu W}$ only.

\subsection{Slow Periodic LS Strategy}
For slow packet rates ($\Tc > 64 \Tsf$) empty sleep frames are automatically sent 
by the LSE on the link transmitter to enlarge the sleeping period without worsening responsiveness.
Power consumption on both sides can be derived from the periodic case by including this additional contribution
\begin{align}
	\Ptx^\mathrm{slP} &= \Ptx^\mathrm{P} + \Etxe \Nempty \Lambdac, \\
	\Prx^\mathrm{slP} &=\Prx^\mathrm{P} + \Erxe \Nempty \Lambdac,
\end{align}
where $\Nempty = \left\lceil \tauc/64 \right\rceil -1$ is the number of empty sleep frames inserted after every data frame.
Because of our assumptions, $\Nempty \Lambdac < \Lambdasf / 64 $, which sets an upper bound to the additional consumption needed to keep LS alive.

\subsection{Extended Periodic LS Strategy}
We have shown that when the link conveys both sporadic streams of packets with relative deadline $\Tmin$ and a periodic stream with period $\Tc$ the best approach is to adopt \textit{xsleep} commands where $\Nsleep = \lfloor \tauc \rfloor -1$ and $\Nsnooze = \lfloor \taud \rfloor -1$.
With this strategy, the receiver wakes up temporarily  $\Nwup$ times before listening is re-enabled, 
and $\Nwup = \lceil (\Nsleep+1) / (\Nsnooze+1) \rceil -1 = \left\lceil {\left\lfloor \tauc \right\rfloor}/{\left\lfloor \taud \right\rfloor}\right\rceil \!-\!1$.
For example, if $\Tc=\unit[600]{s}$ and $\Tmin=\unit[30]{s}$,
then $\Nsleep=296$, $\Nsnooze=13$, and the link is awakened $21$ times during every sleeping period.
For the sake of simplicity, we do not consider cases where $\Tc>4096\Tsf$.

Power consumption on the transmitting side is almost the same as for the periodic strategy
\begin{align}
	\Ptx^\mathrm{xP} = \Ptx + \Lxsleep \EBtxd \Lambdac,
\end{align}
however the size $\Lxsleep$ of the sleep IE is slightly larger than the basic 
sleep command ($\Lxsleep=\unit[5]{B}$).

On the receiving side
\begin{align}
	\Prx^\mathrm{xP} 	\nonumber
	&= \Prz + \Lxsleep \EBrxd \Lambdac + \Eil \left[ \Lambdasf - \left( 
	\lfloor \tauc \rfloor - \Nwup \right) \Lambdac \right] \\ 
	&= \Prx^\mathrm{P} \!+\! (\Lxsleep \!-\! \Lsleep) \EBrxd \Lambdac
	\!+\! \Eil \Nwup \Lambdac.
\end{align}
As expected, power consumption is higher than for basic sleep commands with the same value of $\Nsleep$, because of the additional contribution of idle listening introduced by the awakening mechanism to ensure better responsiveness.

\section{Performance Comparison}
\label{sec:PC}
To compare strategies we assume that both periodicity and relative deadlines are known (or correctly estimated by PRIL).
The metric we consider is the power required to deal with a packet stream that transfers application data on a single link. Only contributions due to communication are taken into account,
as other kinds of power drain by motes can be reduced considerably by exploiting careful hardware and software optimization.
The frame size $\LB$ is selected equal to $\unit[90]{B}$,
to resemble ICMP echo requests generated by \texttt{ping} commands in real networks.
Three different packet transmission rates $\Lambdac$ were considered, 
corresponding to periods equal to $\unit[30]{s}$, 2 minutes and 10 minutes respectively, while 
$\Etxe=\unit[87]{\mu J}$ and $\Erxe=\unit[117]{\mu J}$ for the slow periodic LS strategy (labeled  ``Basic (slow)'' in Table~\ref{tab:res}). 
These values were obtained by considering empty sleep frames encoded with $\unit[40]{B}$.

Results are reported in Table~\ref{tab:res}.
TSCH shows noticeably higher consumption than the oracle on the receiving side.
Tangible saving is possible with the periodic LS strategy based on sleep commands (``Basic'')
and improves with larger periods.
Unfortunately, responsiveness is also impaired, and this could be a problem if sporadic packets were occasionally conveyed on the same link.
The periodic LS strategy based on extended sleep commands (``eXtended'') is a reasonable trade-off and is suited to those cases where the period is much larger than the relative deadline.
By means of two degrees of freedom (sleep and snooze times),
it achieves interesting power saving yet keeping the MAC complexity at a minimum.
In general, it retains the expected low-power consumption of TSCH on the transmitting side,
whereas responsiveness and energy saving on the receiver side
can be balanced on a per-link basis.

\begin{table}[t]
    \caption{Power consumption on both sides of the link and worst-case transmission latency $\Twc$
    for the different LS strategies by varying packet generation period $\Tc$ and relative deadline $\Tmin$.}
    \label{tab:res}
    \begin{center}
    \tabcolsep=0.15cm
\begin{tabular}{|c|c|c|c|c|c|c|c|}
	\hline
	$\Tc$ / $\Tmin$ &  Strategy	& $\Nsleep$ & $\Nsnooze$ 	& $\Twc$ 	& 	$\Ptx$ 		& $\Prx$ 	\\
	($\unit[]{s}$) / ($\unit[]{s}$) &  & & & ($\unit[]{s}$) & 	($\unit[]{\mu W}$)  & ($\unit[]{\mu W}$) \\
	\hline \hline
	
	$30$ / -- & Oracle 		    & -- 		& -- 		& $2.02$	& $8.8667$ 	& $9.6000$ \\ \hline
	$30$ / -- & TSCH 		    & -- 		& -- 		& $2.02$ 	& $8.8667$ 	& $73.3168$ \\ \hline
	$30$ / -- & Basic 		    & 13 		& -- 		& $28.28$   & $9.0667$ 	& $13.6468$ \\ \hline \hline
	
	$120$ / -- & Oracle 		& -- 		& -- 		& $2.02$	& $2.2167$	& $2.4000$ \\ \hline
	$120$ / -- & TSCH 			& -- 		& --    	& $2.02$ 	& $2.2167$	& $69.5668$ \\ \hline
	$120$ / -- & Basic 		    & 58 		& -- 		& $119.18$ 	& $2.2667$ 	& $2.8993$ 	\\ \hline
	$120$ / $10$ & eXtended 	& 58 		& 3			& $8.08$ 	& $2.3000$ 	& $19.0210$ \\ \hline
	$120$  / $30$ & eXtended 	& 58 		& 13		& $28.28$ 	& $2.3000$ 	& $7.5210$ 	\\ \hline \hline
	
	$600$ / -- & Oracle 		& -- 		& -- 		& $2.02$    & $0.4433$ 	& $0.4800$ \\ \hline
	$600$ / -- & TSCH 			& -- 		& -- 		& $2.02$ 	& $0.4433$ 	& $68.5668$ \\ \hline
	$600$ / -- & Basic (slow)   & 296 		& -- 		& $129.28$ & $1.0333$  & $1.2733$  \\ \hline	
	$600$ / $10$ & eXtended 	& 296 		& 3		    & $8.08$     & $0.4600$	& $17.5177$ \\ \hline	
	$600$ / $30$ & eXtended 	& 296 		& 13		& $28.28$ 	& $0.4600$	& $5.3277$ \\ \hline
	$600$ / $120$ & eXtended 	& 296 		& 58		& $119.18$ 	& $0.4600$	& $1.6477$ \\ \hline
\end{tabular}
	\end{center}
\end{table}

\section{Conclusion}
\label{sec:CONC}
By layering time slotting and channel hopping atop IEEE 802.15.4, TSCH offers high reliability, determinism, and low power consumption, which makes it a good candidate when mesh WSNs compliant to the IIoT paradigm have to be deployed in industrial scenarios.
Unfortunately, the access technique of TSCH (but the same also holds for similar protocols that rely on scheduled access) implies non-negligible power consumption by receivers due to idle listening,
i.e., when the receiving circuitry is switched on but no one is sending frames on air.

Details about the PRIL proposal have been presented in this paper, in order to reduce idle listening phenomena by exploiting available information about data exchanges.
In particular, we focused on the listening suspension (LS) mechanism, to be incorporated in the TSCH MAC, that enables the receiving side of a link to be put on sleep on demand.
Besides basic strategies, suitable for periodic packet streams, extended solutions have also been introduced, 
which offer a reasonable trade-off between power consumption and responsiveness for sporadically generated packets thanks to the ability to repeatedly snooze the receiver.

Formulas have been derived for power consumption estimation for both sides of a link in 6TiSCH,
which were used to evaluate the behavior of a link in realistic operating conditions, by leveraging the energy model of a real device (OpenMote B running OpenWSN).
Results confirm that the amount of saved energy is worth the adoption of LS.
Future activities in this area will focus on the evaluation of LS mechanisms through simulation of realistic (and more complex) traffic patterns and the implementation of PRIL algorithms to automatically infer fitting traffic models.
 
\bibliographystyle{IEEEtran}
\bibliography{bibliography}

\end{document}